\documentstyle[astrobib]{mnv2}

\input psfig

\title[Distribution functions for galaxy cluster models]
{Distribution functions for clusters of galaxies \\ from N-body simulations}

\author[Priyamvada Natarajan, Jens Hjorth and Eelco van Kampen]
  {Priyamvada Natarajan$^{1}$, Jens Hjorth$^{1}$ and Eelco van Kampen$^{2}$ \\
$^{1}$Institute of Astronomy, Madingley Road, Cambridge CB3 0HA \\
$^{2}$Royal Observatory, Blackford Hill, Edinburgh  EH9 3HJ\\}

\begin{document}
\label{firstpage}
\maketitle

\begin{abstract}
We present the results of an attempt to adapt the distribution
function formalism to characterize large-scale structures like clusters of
galaxies that form in a cosmological N-body simulation. While galaxy
clusters are systems that are not strictly in equilibrium, we show that
their evolution can nevertheless be studied using a physically motivated 
extension of the language of equilibrium stellar dynamics. Restricting
our analysis to the virialized region, a prescription to limit the
accessible phase-space is presented, which permits the
construction of both the isotropic and the anisotropic distribution
functions $f({\cal E})$ and $f({\cal E}, L)$. The method
is applied to models extracted from a catalogue of simulated clusters. 
Clusters evolved in open and flat background cosmologies are 
followed during the course of their evolution, and are found to transit
through a sequence of what we define as `quasi-equilibrium' 
states. An interesting feature is that the computed $f({\cal E})$ 
is well fit by an exponential form. We conclude that the dynamical
evolution of a cluster, undergoing relaxation punctuated by 
interactions and violent mergers with consequent energy-exchange,
can be studied both in a qualitative and quantitative fashion by
following the time evolution of $f({\cal E})$.
\end{abstract}


\section{Introduction and Motivation}

In this paper we present the results of an attempt to construct
distribution functions for clusters of galaxies that form in N-body
simulations. To permit the description of systems like clusters 
that are continually evolving at the present epoch
we propose a possible extension of the distribution function
formalism of equilibrium stellar dynamics. The prescription developed
is applied to simulated clusters for which the isotropic and
anisotropic distribution functions are computed. We then use the 
time evolution of the distribution function to get an insight into 
the details of the relaxation process and energy exchange in clusters.

The motivation for this work is twofold: (i) to develop
a mathematical formalism to characterize the dynamical evolution
of `stellar' systems that are marginally out of equilibrium, which 
can be applied to numerical cluster models and (ii) to gain insight
into the dynamical evolution and properties of clusters in general with
as special emphasis on the origin of density profiles and the use of 
clusters as cosmological probes.

Previous studies that are relevant to our work have been done in
three somewhat distinct contexts: (i) studies of the dynamical state
of clusters formed in N-body simulations, (ii) studies of the
formation and evolution of elliptical galaxies in an attempt to
understand the physical origin of the universal $R^{1/4}$ law
and (iii) the evolution of stellar systems studied in phase space. 

Given the standard paradigm of Cold Dark Matter (CDM) dominated
hierarchical structure formation scenarios, the formation, internal
structure and evolution of dark halos and density profiles has been
studied in detail for a range of initial input power spectra evolved
in a slew of background cosmological models
(\citeNP{fillmore84}; \citeNP{hoffman85}; \citeNP{west87};
\citeNP{crone94}; \citeNP{richstone92}; \citeNP{cen93}; \citeNP{klypin};
\citeNP{cen94}; \citeNP{navarro94a}; \citeNP{dodds95}).
A `universal density profile' has been found to be a good fit over a 
two-decade range of scales for dark halos that form in N-body realizations
of the cosmological model (\citeNP{navarro96}; \citeNP{simon96}).
These authors and several other groups 
(\citeNP{lacey96}; \citeNP{tormen96}) find that for a range of
cluster-halo shapes, sizes and other bulk properties, the profile  
depends primarily on the epoch of formation and only rather weakly on the
details of the cosmological context and the initial power spectrum of 
fluctuations. 
\citeN{navarro96} find that their 
spherically averaged density
profiles can be fit by scaling a `universal' profile (henceforth
referred to as the NFW profile) which goes as $\rho\,\propto\,{r^{-1}}$ at 
small radii and steepens to $\rho\,\propto\,{r^{-3}}$ at large radii.
Subsequently, \citeN{lacey96} on examining the internal structure 
of dark matter halos formed in scale-free hierarchical models
for a range of initial power spectra, also find that the density
profile is well-fit by both the
NFW profile and the analytic \citeN{hernquist90a} profile which falls
off as $\rho\,\propto\,{r^{-4}}$ at large radii. Part of the aim of
the present investigation is to explore
the interplay between the initial conditions and subsequent relaxation
processes in clusters and to understand the possible physical origin 
of these density profiles. 

This approach is somewhat analogous to the 
detailed studies of the origin of the $R^{1/4}$ law in elliptical 
galaxies (\citeNP{vanalbada82}; \citeNP{carlberg86}; \citeNP{londrillo91}; 
\citeNP{hjorth91}). Moving over to the analogy with the formation of 
elliptical galaxies, we briefly review some scenarios that have been 
put forward to understand the observed universal $R^{1/4}$ profile.
The physical system in this case essentially consists of stellar orbits 
in a potential which deepens during the course of evolution via violent
relaxation that occurs as a consequence of strong collective potential
fluctuations (\citeNP{dlb67}). During the course of evolution 
following the violent phase, any remaining positive energy particles 
slowly diffuse out and eventually leave the system. 
In the case where the resultant stellar system has a deep potential
well the $R^{1/4}$ surface-brightness profile arises as a natural
consequence (\citeNP{hjorth91}).
Such a final configuration can be produced from a cold dissipationless 
collapse scenario leading to stellar orbits that are preferentially radially 
anisotropic as well as from warmer initial conditions with 
dissipation.  

While universality intuitively suggests
wiping out of the memory of initial conditions and similarities in the
post-collapse evolution, \citeN{londrillo91} conclude from their
study of dissipationless galaxy formation that the 
phase-space properties of the final equilibrium state do in fact depend
strongly on the details of the initial conditions. They find that the 
relaxation to an $R^{1/4}$ profile is characterized by two phases: 
(i) an early phase dominated by strong potential and density
fluctuations in the central region which results in a core-halo
structure and a long tail in both the
density and energy distribution and (ii) a longer-lived later phase
wherein the system evolves mainly due to the effect of small-scale
diffusion which smoothes the connection between the core and the
halo populations. Some memory of the early phase is retained, since the 
fluctuating forces and the compact core necessary to diffuse the 
low angular momentum orbits in the later phase are necessarily
remnants of the initial collapse dynamics; i.e. the region of phase
space that is available for diffusion into for the system during the
late phase is delineated during the early phase. Also in
dissipational scenarios (\citeNP{barnes96}) which are expected to introduce
irreversibility, especially during mergers, some memory of the initial
conditions seems to be retained. 

Finally, we place this work in the context of studies of the evolution
of stellar systems in phase space. The structure of phase space and
its accessibility for $R^{1/4}$ galaxies was explored by
\citeN{binney82a} who found that the distribution of the number
of stars of a given energy is approximated rather well by an
exponential formula. It was argued that if elliptical galaxies formed 
via dissipationless processes, the exponential form found at late
times for the distribution function was probably set up during 
the epoch of formation, yielding a scale-free form for the density
profile. Therefore, the phase space corresponding to 
the observed surface brightness profiles of elliptical galaxies are 
consistent with having been demarcated during the initial collapse 
of a density fluctuation that subsequently gives rise to the galaxy.   
In an another study, \citeN{hernquist93} attempted to quantify 
the importance of mergers in the origin and structure of early-type 
galaxies by exploring the phase-space properties of merger remnants 
and comparing them to simple models of elliptical galaxies.   

In a more recent analysis, \citeN{voglis} presents an analytic model
fit to the distribution
function of a nearly spherical, anisotropic galaxy-scale system. 
Starting with cosmologically consistent initial conditions, he finds 
a dichotomy of states, a core population dominated by low
angular-momentum tightly bound particles and a less bound halo population.     
The energy distribution of the core population is nearly isothermal
whereas the halo population consists of particles on radial
orbits with correlated energy and angular momentum. Following the time
evolution of a radius containing a specific fraction of the mass, he
finds that the radial oscillations decay quickly after collapse
preceding the onset of relaxation during which a new equilibrium is
established. 

Despite some essential physical differences between
stellar systems and clusters, we synthesize the approaches described
above and demonstrate that they can be adapted and successfully
applied to the case of galaxy clusters that form in realizations of 
cosmological N-body simulations. Specifically, we develop in the
present paper a  framework that can be applied to characterizing what we shall
define as {\it quasi-equilibrium
evolution}. In Section 2 we outline our conceptual model
of a cluster, review the equilibrium distribution function formalism
and present our prescription to extend it to quasi-equilibrium.
The details of the N-body simulations from which 
the cluster catalogs that we study were extracted are described in 
Section 3. In Section 4 we apply the method to the construction 
of $f({\cal E})$ and $f({\cal E},L)$ for slices from the simulations. 
We present our results and summarize what can be learnt from a
first set of
diagnostics developed to describe the quasi-equilibrium
evolutionary states of the cluster in Section 5.

\section{THE DISTRIBUTION FUNCTION FORMALISM}
 
\subsection{Characterizing a cluster}

A cluster is a composite system with diverse components:
collisionless dark-matter particles, collisional baryonic gas and bound
stellar systems (the galaxies).
The rudimentary physical picture of the cluster that we find useful, 
is one of a system with two natural length scales delineated by
the dynamics: the virial radius $r_{\rm vir}$, defined to be
the radius enclosing an overdensity of $\approx$ 180 times the
critical density, which is motivated by the collapse of 
the simple spherical top-hat model (\citeNP{peebles80}) and the
turnaround radius $r_{\rm ta}$, defined to be the radius at which the
cluster separates from the general cosmological
expansion (envelope with zero
relative velocity). There is on-going infall inward
of $r_{\rm ta}$, and therefore the system can never really be
treated as one evolving in isolation. 

It is instructive to clarify at this juncture some of the terms and 
their definitions that will be employed in our analysis to describe the
dynamical state of the system. A {\it{virialized}} system is one for
which the virial theorem holds (${{d^2{I_{ij}}}/{dt^2}}\,=\,0$; where
${I_{ij}}$ is the moment of inertia tensor) and therefore there exist 
no systematic motions, neither expansion nor contraction; a 
{\it{stationary}} system is one for which there is no time variation 
of the distribution function $f$, 
i.e. $\partial f/\partial t\,=\,0$, and a {\it{relaxed system}} is one for
which memory of the initial conditions in phase-space has been erased.
For a cluster-scale collisionless system, two-body relaxation processes
are unimportant since the time-scale relevant for two-body processes
far exceeds the dynamical time, although in the case of a simulated
cluster, there arise some purely numerical two-body effects which we discuss in 
more detail in Section 3. While collective energy exchange processes 
might occur during the course of dynamical evolution, they might not
always lead to a relaxed final state and also a virialized system need
not be relaxed. This is so because virialization is defined in terms
of the kinetic and potential energies of a specific configuration of the system
at a given epoch, while relaxation requires physical processes that affect the
evolution to be taken into account and is hence a statement about the 
integrated dynamical history. 

In this study we focus on the collisionless component.
We define a collisionless equilibrium system to be one that is virialized,
stationary and relaxed with no on-going energy exchange. Clearly, this
is not the case for a cluster of galaxies, wherein
departures from equilibrium necessarily occur due to the existence of
an infall region and departure from stationarity which is a consequence of
fluctuations induced in the potential in response to mergers and
secondary infall. Nevertheless, 
there does seem to exist a virialized central region and 
observed clusters seem not to be far from 
equilibrium as inferred from their `measured' density profiles and 
velocity dispersions. Therefore, the need arises to develop a
phase-space description that can incorporate small departures
from equilibrium, which we attempt below. To first approximation,
we model a cluster in this treatment as a spherically symmetric
and non-rotating system.

\subsection{The equilibrium distribution function formalism}

For a statistical description of a system with a large number of 
particles, it is convenient to define the distribution
function (DF) (the phase-space mass density) $f(\bmath r, \bmath v, t)$.
A given configuration of the system is then specified by $f(\bmath
r,\, \bmath v,\, t ) \, {\rm d}^{3} \bmath x \, {\rm d}^{3} \bmath v$ -- the
mass of particles having positions in the infinitesimal volume 
element ${\rm d}^{3} \bmath r$, with velocities in the range
${\rm d}^{3} \bmath v$ in the $({\bmath r},{\bmath v})$ phase space. The DF
satisfies the Boltzmann equation,
\begin{equation}
{\frac {{\rm d} f}{{\rm d} t}} = {\frac {\partial f}{\partial t}} + {\bmath
v}\,{\cdot}{\bmath \nabla} f - {\bmath \nabla \Phi}{\cdot}{\frac
{\partial f}{\partial \bmath v}} = C, 
\end{equation}
where ${{\rm d} f}/{{\rm d} t}$ is the total derivative along the exact
trajectory of the particle moving in the force field ${\bmath \nabla \Phi}$
and $C$ is the collisional term, which defines the changes in $f$ due to
collisions between particles. In conjunction with the Poisson equation,
\begin{equation}
{\bmath \nabla^{2}}\,\Phi\,=\,4 \pi G \rho,
\end{equation}
and the definition of the density profile,
\begin{equation}
\rho({\bmath r}, t)\,=\,\int\,f({\bmath r},{\bmath v}, t)\,{{\rm d}^{3}}{\bmath v},
\end{equation}
equations (1), (2) and (3) constitute a complete set of
self-consistent evolution equations that describe a self-gravitating
system. For a stellar system that is strictly  collisionless and
gravitating, $C$ = 0 in equation (1), the solutions to the above
are the stationary states, and the particle trajectories are the 
`characteristics'. Additionally, while equation (2) is valid only for a self-gravitating system,
the potential defined in equation (1) could also include the contribution
from an external potential.

We review below some of the basic definitions that will be
used throughout the paper; a more comprehensive treatment of the
formalism can be found in \citeN{b&t87}. The notational conventions
followed here are also as in \citeN{b&t87} except for the introduction
of the function $h$. 

The binding energy per unit mass ${\cal E}$ is defined as,
\begin{eqnarray}
{\cal E}\,=\,{\Psi(r)}\,-\,{{v^2} \over 2},
\end{eqnarray}
and the angular momentum per unit mass $L$ is given by,
\begin{eqnarray}
L\,=\,|{\bf r}\,\times\,{\bf v}|.
\end{eqnarray}
The DF for a spherically symmetric, non-rotating stellar system can
only depend on these two quantities. Dependence on both ${\cal E}$ and
$L$ gives rise to an anisotropic velocity distribution whereas
dependence on ${\cal E}$ only yields an isotropic distribution of
velocities.
The differential energy distribution
$({{{\rm d}M({\cal E})}/{\rm d}{\cal E}})\,{\rm d}{\cal E}$ 
is the total mass of the system with binding energy within
$[\,{\cal E},\,{\cal E}+{\rm d}{\cal E}\,]$
which is then explicitly defined to be,
\begin{eqnarray}
h({\cal E})\,{\rm d}{\cal E}\,\equiv\,{{{\rm d}M({\cal E})} \over
  {\rm d}{\cal E}}\,{\rm d}{\cal E}\,=\,\int_{\Delta V({\cal E})}\, 
  f({\cal E})\,{{\rm d}^{3}}{\bmath r}\,{{\rm d}^{3}}{\bmath v},
\end{eqnarray}
integrated over the volume element ${\Delta V({\cal E})}$ in phase-space with
energy in the specified range $[\,{\cal E},\,{\cal E}+{\rm d}{\cal E}\,]$.
Since we are dealing with spherically symmetric systems we can write this
integral in terms of $r$ and $v$. Furthermore, we can take $f({\cal E})$
out of the integral and use eq.\ (4) to arrive at:
\begin{eqnarray}
h({\cal E})\,{\rm d}{\cal E}\, = \,16 {\pi^2}f({\cal E})\,{\rm d}{\cal E}
\int_0^{r_{\rm m}({\cal E})}{r^2}\,{\rm d}r 
\sqrt{2\,(\Psi(r)-{\cal E})}.
\end{eqnarray}
This can be written as
\begin{eqnarray}
h({\cal E})\,{\rm d}{\cal E}\,=\,f({\cal E})\,g({\cal E})\,{\rm d}{\cal E},
\end{eqnarray}
where $g({\cal E})$ the density of available states in phase space of
energy ${\cal E}$,
\begin{eqnarray}
g({\cal E})\,\equiv\,{16 {\pi^2}}{{\int_{0}^{r_{\rm m}(\cal E)}}{\sqrt{2\,(\Psi(r) - {\cal E})}}\,{r^2}\,{\rm d}r}.
\end{eqnarray}
This integral is strictly defined only for $\Psi(r)\,\geq\,{\cal E}$.
Here ${r_{\rm m}(\cal E)}$ is defined to be the radius where
$\Psi({r_{\rm m}})\,=\,{\cal E}$.
Thus, we obtain an operational definition for $f({\cal E})$,
\begin{eqnarray}
f({\cal E})\,=\,{h({\cal E}) \over g({\cal E})}.  
\end{eqnarray} 
Given a potential $\Psi(r)$, both $h({\cal E})$ and $g({\cal E})$ are
well-defined for stellar systems in equilibrium. 
We use the definition in equation (9) to construct the DF in our
analysis. The DF could
alternatively be derived and hence constructed by either explicitly counting
available and energetically accessible cells in phase space or from the
density profile of the system $\rho(r)$ using an inversion method devised by
\citeN{eddington16}. Quoting the result of Eddington's inversion
formula from \citeN{binney82a},
\begin{eqnarray}
f({\cal E})\,=\,{1 \over {{\sqrt 8}{\pi^2}}}\,{{\int ^{\cal
      E}_{0}}\,{{{{\rm d}^2} \rho} \over {{\rm d}{\Psi^2}}}\,\,{{{\rm d}\Psi} \over
{\sqrt{\Psi\, - \,{{\cal E}}}}}}\,+\,{{1 \over {\sqrt{{\cal
      E}}}}\,{{{\rm d}{\rho}} \over {\rm d{\Psi}}}},
\end{eqnarray}
\begin{eqnarray}
{{{{\rm d}^2} \rho} \over {{\rm d}{\Psi^2}}}\,=\,{\left({r^2 \over
      GM}\right)^2}\,
\left[\,{{{\rm d^2}\rho} \over {{\rm d} r^2}} \,+\,\,{{{\rm d} \rho} \over {{\rm d} r}}\left
({2 \over r}\,-\,{{4 \pi \rho {r^2}} \over M(r)}\right)\right]\,
\end{eqnarray}
where $M(r)$ is the total mass enclosed within radius $r$ of the system. 

For a spherically symmetric system with an anisotropic velocity
distribution, the DF $f({\cal E},\,L)$ is analogously defined to be,
\begin{eqnarray}
f({\cal E},\,L)\,=\,{h({\cal E},\,L) \over g({\cal E},\,L)}.  
\end{eqnarray}
The density of states in phase-space $g({\cal E},\,L)$ is now a hypersurface of energy
${\cal E}$ and angular momentum $L$. Furthermore,
\begin{eqnarray}
g({\cal E},\,L)\,=\,{8 {\pi^2}}\,L\,{T_{\rm r}({\cal E},\,L)},
\end{eqnarray}
where ${T_{\rm r}}$ is the radial period defined as,
\begin{eqnarray}
{T_{\rm r}({\cal E},\,L)}\,=\,{\int_{r_{\rm 1}}^{r_{\rm 2}}}\,{{{\rm
d}r} \over {\sqrt{2(\Psi(r) - {\cal E})\,-\,{{L^2} \over {r^2}}}}},
\end{eqnarray}
the range of integration being from the pericenter of the orbit,
${r_{\rm 1}}$, to the apocenter, ${r_{\rm 2}}$. The radial period
${T_{\rm r}}({\cal E},\,L)$
is generally evaluated numerically, but an approximately analytic 
expression can be obtained for a few special choices of the potential; 
one of them being the isochrone potential, for which the following
simple expression holds, 
\begin{eqnarray}
{T_{\rm r}}({\cal E},\,L)\,=\,{{2 \pi G M} \over {(2 {\cal E})^{3/2}}}.
\end{eqnarray}
Furthermore, $T_{\rm r}({\cal E},L)$ has been demonstrated
to be almost independent of $L$ for different potentials (isochrone 
and Keplerian in particular) by \citeN{voglis} and \citeN{lynden-bell96}.

\subsection{Adapting the DF formalism to quasi-equilibrium}

The equilibrium formalism outlined above does not strictly apply to 
clusters of galaxies as these systems are constantly evolving. 
Additionally, the equilibrium formalism as currently posed 
cannot accommodate the existence of positive energy particles that 
are sometimes observed in numerical cluster simulations (see plots in
Section 5). On the other hand, it appears to be sensible to attempt using the equilibrium
formalism as a guide to probe the non-equilibrium domain.
Indeed, most studies of the dynamics of clusters of galaxies implicitly
assume that an equilibrium description is an adequate starting point. 

Below we first describe the inherent limitations that such an approach
is necessarily subject to, but we
then use precisely these restrictions to formulate a physically
motivated and self-consistent extension of the equilibrium formalism. The 
key to the extension 
arises from the recognition of the important distinguishing 
features in the case of clusters: the finite time available for evolution
and the spatial extent of a cluster.

Examining the relevant time-scales: firstly, the typical
time that it takes for a cluster to be virialized out to the
Abell radius is of the order of the Hubble time which in itself 
precludes the cluster from reaching an equilibrium configuration.
Secondly, the disturbances from continuous infall and the more violent
interactions constantly drive the system away from equilibrium whereas
the collective energy exchange processes tend to drive the system 
toward equilibrium. The fact that the characteristic time-scale on
which the collective energy exchange process operate is relatively 
short means that the system is capable of quickly recouping from or 
adapting to a disturbance. Thus, it is possible that the system might 
come close to equilibrium fairly quickly after such
events. In other words, it may be possible to decouple the system into 
two components: a fast component and a slowly evolving component, akin
to the adiabatic approximation.
We may therefore view the system as evolving from one 
{\it quasi-equilibrium}\/ state to another on a time scale 
that is roughly the `disturbance time scale'. 

Having thus argued that 
the system might frequently be close to a quasi-equilibrium state, we 
need to delineate more precisely the specific region to which such a 
treatment is most suitably applicable. Since the use of the
equilibrium formalism necessarily requires the system to be
virialized, a natural choice would be to study a typical virialized 
region; clearly, the infall region cannot be studied via this
approach. Also, once a radially infalling 
particle has entered the virialized region its kinetic energy gets 
distributed to other particles and it is more likely to get bound 
inside the virialized region rather than leave the system. Moreover,
even the few particles that are capable of leaving the region
are more than balanced in the net by the infalling ones.

In summary, the finite time and extent arguments coupled with the expected
short relaxation time scale encourage us to develop the notion of
quasi-equilibrium states further. However, the eventual usefulness and
indeed validity of the approach should be judged not on these
arguments alone but on the results presented in Sections 4 and 5.
We reiterate that our proposed adapted equilibrium 
formalism is only applicable to a restricted region in phase space.

There have been several previous studies that have attempted to limit 
the available phase space for elliptical galaxies
(\citeNP{stiavelli85}; \citeNP{tremaine87}; \citeNP{merritt89}).
For instance, \citeN{spergel92} suggested a prescription to limit the
macroscopic states that are accessible to individual particles in a
collisionless system approaching equilibrium by approximating violent
dynamical processes en-route to relaxation as a sequence of discrete
scattering events. They also introduced a monotonically increasing
non-equilibrium entropy by imposing additional constraints on the
available phase-space. 

In our approach we make no {\it a priori}\/ assumptions about the nature
of energy-exchange processes and do not attempt to define any
generalized entropy for the system. Rather we shall argue that it is the
usual definitions (8) and (13) of $g({\cal E})$ and $g({\cal E},L)$
that need to be modified in a consistent fashion to enable the
description of systems evolving marginally away from equilibrium. 

\subsubsection{Computing g(${\cal E}$) for non-equilibrium systems}

As described in Section 2.2, the phase-space density $g({\cal E})$
is valid only for negative energies (positive ${\cal E}$, i.e.\ strictly
bound particles in an equilibrium system), and diverges for ${\cal
  E}\rightarrow 0$ due to the
blowing up of the volume element in physical space. For instance,
consider a potential $\Psi(r)$ with a radial dependence such that,
\begin{eqnarray}
\Psi(r)\,=\,{r^{-\alpha}};\ \ \ \ 0< \alpha \le 1.
\end{eqnarray}
In this case
${r_{\rm m}}({\cal E})\,\sim\,{{\cal E}^{-\alpha}}$ and so
$g({\cal E})$ diverges as ${\cal E}^\eta$ for 
$\eta=\alpha^2/2-3\alpha$, i.e., $-5/2\le\eta < 0$.
Requiring $f({\cal E})$ to be finite at the escape energy and the mass
to be finite simultaneously implies that $\eta\,>\,-1$, in turn
restricting $0\,<\,\alpha\,<{3-\sqrt5}$ which for the
self-gravitating case $(\alpha\,=\,1)$ means that the asymptotic behavior 
of $g({\cal E})$ cannot be compensated by a similar divergence of 
$h({\cal E})$. 
Furthermore, $g({\cal E})$ is infinite for all ${\cal E} \le 0$, as
the accessible phase-space is infinite for these energies. 
As discussed above, we propose to limit the accessible phase-space due to the
finite time available. We do this in practice by 
truncating $g({\cal E})$ at a fixed scale $r_{\rm max}$. 
As we show below, this will remove the divergence and 
render $g({\cal E})$ well-behaved for all ${\cal E}$.

Physically, such a truncation
amounts to the assumption that for most of the bound particles that
constitute the system, the finite time elapsed since the initial collapse 
necessarily implies that the probability of occupation of states 
falls off steeply beyond a fiducial volume in phase space which is dictated by 
the threshold distance that a particle with typical velocities can 
traverse in the available time interval. Energy-exchange and hence
marginal departures from equilibrium of the
cluster arise both from the existence of the extended infall region as
well as due to the occurence of frequent mergers and interactions as
an integral part of the relaxation process. Since, in this analysis, 
we are primarily interested in probing and characterizing the
evolution of the central regions and the relaxation effects therein,
we choose to truncate the system at
$r_{\rm max}\,=\,r_{180}\,=\,r_{\rm vir}$, the radius within
which the system is expected to be virialized, and is indeed found to
be for several different power spectra (\citeNP{lacey96}).
This prescription is independent of the functional form for the
potential and can in general be applied to any physically
meaningful potential for a spherically symmetric system. 

Specifically, consider a stellar system consisting of
particles confined in a potential $\Psi(r)$, modelled explicitly as the
Hernquist potential,
\begin{eqnarray}
\Psi(r)\,=\,{{\Psi_{0}} \over {1 + {\frac  {r}{s}}}}.
\end{eqnarray} 
\begin{figure}
\psfig{figure=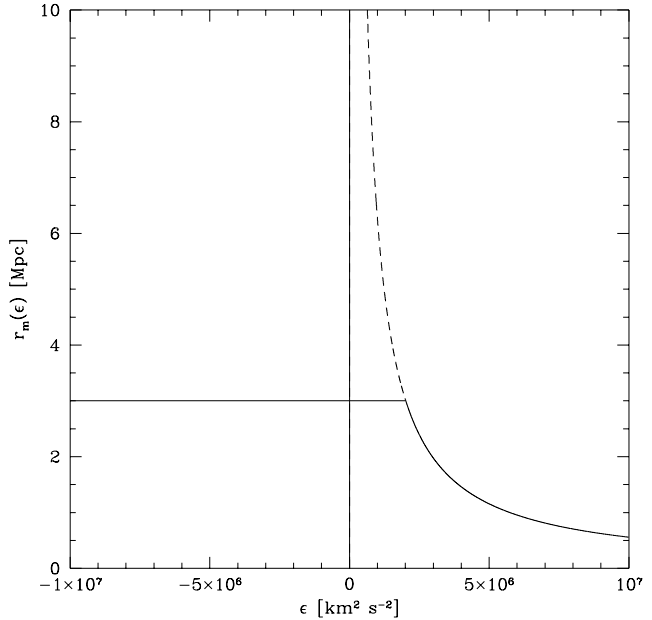,width=0.5\textwidth}
\caption{The prescription for truncation: 
we illustrate the prescription we have used to define $r_{\rm max}$,
the truncation radius, for the Hernquist potential. The solid curve
is our choice for $r_{\rm max}$, which now stands well-defined for
all ${\cal E}$.} 
\end{figure}
In the equilibrium context, the limits of integration for $g({\cal E})$
range from 0 to ${r_{\rm m}({\cal E})}$ defined 
to be the radius at which $\Psi({r_{\rm m}})\,=\,{\cal E}$.

From the plot in Fig.~1, we see clearly that the phase-space volume
element is ill-defined as ${\cal E}\rightarrow {0^{+}}$. By setting 
${r_{\rm max}}$ as the maximal allowed value for $r_m({\cal E})$
in eq.~(8), i.e.,
\begin{equation}
{r_{\rm m}({\cal E})}={\rm Min}\Bigl[
s({{\Psi_{0}} \over {\cal E}}-1),\ r_{\rm max} \Bigr] ,
\end{equation}
we truncate the system in physical space, which automatically 
translates into a restriction of available phase-space. In this
way we recover a well-behaved $g({\cal E})$ that is valid for all
${\cal E}$. 

Thus, using only the assumption that the number of particles within
the boundary $r_{\rm max}$ is roughly constant, we modify the definition
of $g(E)$ to allow for quasi-equilibrium states (including
the temporary existence of positive energy particles).
With this modification, most of the equilibrium formalism then
automatically carries over for spherically symmetric systems.

\subsubsection{Computing g(${\cal E}$, L) for non-equilibrium systems}

The construction of $g({\cal E}, L)$ is analogous to the 
above, but we now need to restrict a two-dimensional hypersurface. The key
observation here is that in addition to the restriction in ${r_{\rm
max}}$, the orbital structure needs to be constrained correspondingly
in a consistent fashion. We proceed as follows, by first assuming that
$g({\cal E}, L)$ is separable,
\begin{eqnarray}
g({\cal E}, L)\,=\,L\,{\tilde{g}}({\cal E}),
\end{eqnarray} 
motivated by the functional form of the expression for $g({\cal E}, L)$
(eq.~13), and the observation of \citeN{voglis} that the radial
period $T_{r}({\cal E},L)$ is only a weak function of $L$. 
A functional form for ${\tilde{g}}({\cal E})$ can then be derived
by integrating $g({\cal E},L)$ over $L$:
\begin{eqnarray}
g({\cal E})\,=\int_{0}^{L_{\rm max}}{L{\tilde{g}}({\cal E})}dL =
{\tilde{g}}({\cal E}) L_{\rm max}^2/2\ .
\end{eqnarray}
Therefore,
\begin{eqnarray}
{\tilde{g}}({\cal E})={2g({\cal E}) \over {L_{\rm max}^2} }\ ,
\end{eqnarray}
and
\begin{eqnarray}
g({\cal E}, L)={{2g({\cal E})L} \over {L_{\rm max}^2} }\ ,
\end{eqnarray}
where the expression is valid for all ${\cal E}$, provided 
$L_{\rm max}$ is appropriately defined as the specific angular
momentum per unit mass for a circular orbit of energy ${\cal E}$. 
Given the prescription for defining $r_{\rm m}({\cal E})$, the maximal
allowed angular momentum for a particle is easily calculated once the circular
velocity at a given radius in the potential is known. For the
Hernquist potential one obtains,
\begin{eqnarray}
L_{\rm max}({\cal E}) = r^2_{\rm m}({\cal E})
{\Psi_0 s \over (1+{{r_{\rm m({\cal E})}}\over s}) }.
\end{eqnarray}

With the modified definitions of $g({\cal E})$ and $g({\cal E},L)$
we have thus successfully adapted the equilibrium DF formalism for
systems in quasi-equilibrium enabling the construction of
both $f({\cal E})$ and $f({\cal E},L)$.

\section{Numerical cluster models}

\subsection{Numerical methods}

We briefly summarize the modelling of the clusters that we intend 
to study. Initial conditions for the cluster
models are generated by means of the
\citeN{rein96} implementation of the Hoffman-Ribak method of constrained
random fields (Hoffman \& Ribak 1991) in order to form a specific cluster
at the centre of the simulation sphere.
Each cluster model is evolved from these constrained initial conditions
by means of the \citeNP{barnes89} treecode, supplemented with a
galaxy formation algorithm (see \citeNP{eelco96a} for details).
During the evolution, galaxies are identified at several epochs and replaced
by single `galaxy particles' in order to ensure their survival, as 
numerical two-body disruption destroys almost all galaxies inside
standard N-body cluster simulations (\citeNP{eelcofirst}; 
\citeNP{carlberg95}). 
Two-component models consisting of distributions of dark-matter
background particles and galaxy particles are thus produced.
A direct match of the simulated galaxy distribution with the
observed one sets the amplitude of the initial density fluctuation
spectrum, and thus the present time in the models. 
The specific prescription
used to define galaxies is necessarily somewhat arbitrary. However,
we consider that it provides at least a crude representation of the
influence galaxies have on the dark matter distribution.
The implementation of alternate prescriptions to define
`galaxies' is unlikely to substantially modify the results presented
here. In any case, for this paper we have examined only the distribution
and dynamics of the dominant dark component.

\begin{figure*}
\vspace{1.5cm}
\psfig{figure=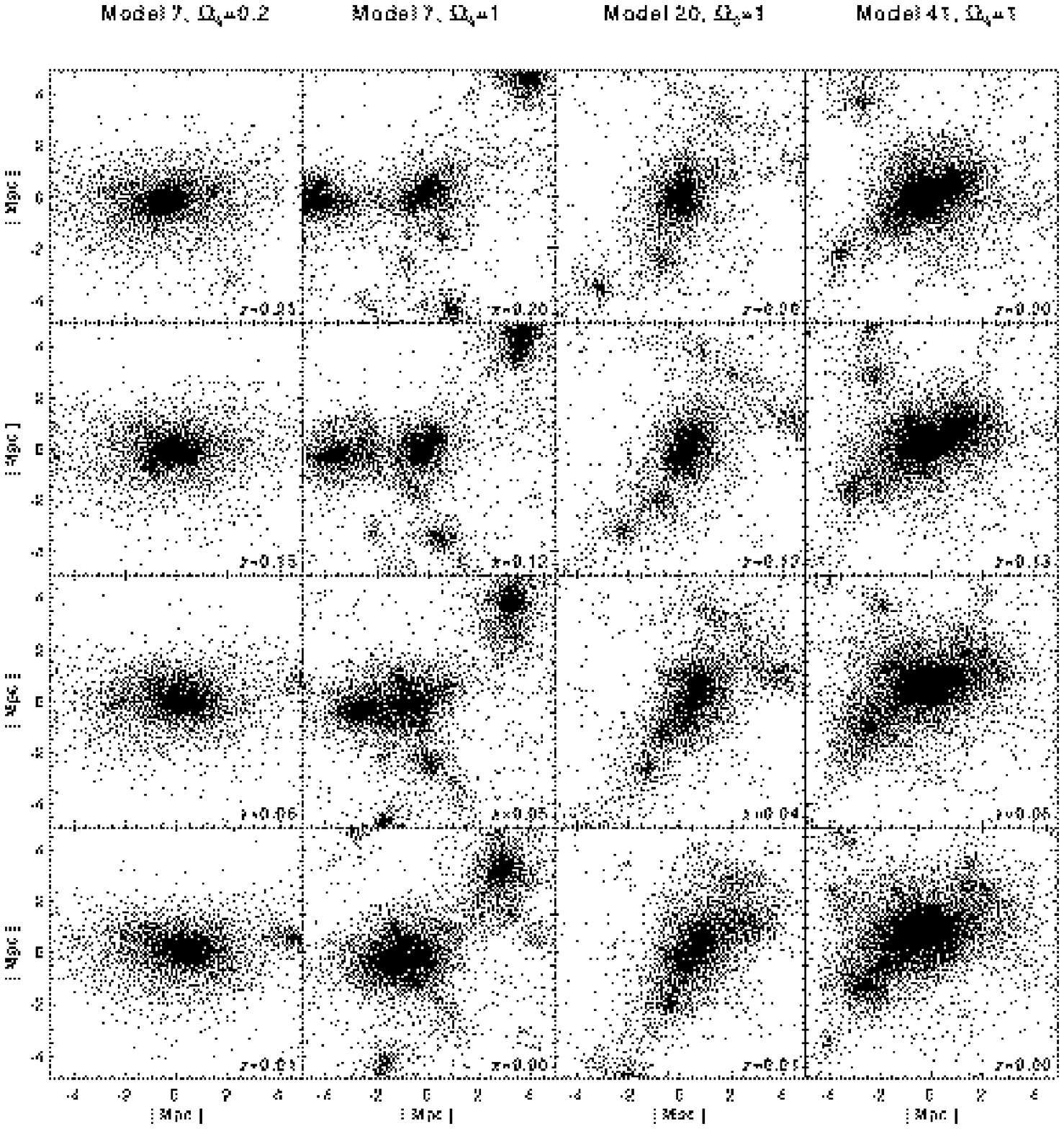,width=18cm}
\vspace{1.5cm}
\caption{The simulation studied here consists of three
cluster models grown in an $\Omega_0=1$ CDM cosmology with
$\sigma_8=0.5$ (the three right columns), and one run of Model 7 with 
$\Omega_0=0.2$; $\sigma_8=1$ (left column), for which identical random 
numbers were used to generate the initial conditions. The slices were
taken so that the interval between the snapshots corresponds to equal
amounts of elapsed time ($0.08t_0$ which is roughly $1\times{10^9}$
yr). For the $\Omega_0=1$ models only half the dark matter particles 
are plotted in these panels, and a tenth of all particles for 
the ${\Omega_0}=0.2$ case.}
\end{figure*}

\subsection{Four selected model clusters}

We use cluster simulations from a model cluster catalogue constructed
by \citeN{katgert96} for the $\Omega_0=1$ standard CDM scenario.
They found a consistent value for $\sigma_8$ (the {\it r.m.s.}\ density
fluctuation within spheres of $8h^{-1}$Mpc) which is used to normalise
the density fluctuation spectrum. A match between models
and data for the galaxy autocorrelation function, the richness
distribution function, and the line-of-sight velocity dispersion
favours $\sigma_8\approx0.4-0.5$. This is consistent with recent results
of \citeN{eke96}, who find $\sigma_8\approx0.5$ for the same scenario.
We therefore adopt a value of 0.5 for $\sigma_8$. The Hubble parameter
$H_0$ is set to 50 km s$^{-1}$ Mpc$^{-1}$ throughout this paper.

For this paper we selected cluster Models 7, 20 and 41 from the full
catalogue. We also built the corresponding models for an $\Omega_0=0.2$
CDM scenario, i.e.\ we kept the same random numbers for the initial
conditions but just changed $\Omega_0$. This low-$\Omega_0$ variant
of CDM was normalized according to the results of \citeN{eke96},
who find from the number density of clusters that $\sigma_8$ scales
with $\Omega_0$ as $0.5\Omega_0^{-0.47+0.10\Omega_0}$.
Therefore, for $\Omega_0=0.2$ we are required to set $\sigma_8\approx 1$,
i.e. to the unbiased value. Here we only consider the $\Omega_0=0.2$ 
version of Model 7. These four models, with around $10^4$ particles
per cluster, form our basic simulation set. We re-simulated all models
from $z=0.2$ up to the present epoch, and saved the particle data
frequently to get a finer timeslicing than the original
run. Furthermore, since we do not expect to
see any new galaxies forming within the clusters at these low redshifts,
we switched off the galaxy formation algorithm in order to simplify the
book-keeping of particle numbers.
In Fig.\ 2 we show four epochs of the time evolution of the selected four
cluster models. This figure will be discussed in more detail in Section 5.1.

\subsection{Resolution and two-body relaxation}

The resolution of an N-body model is primarily set by the particle
softening that has to be employed in order to reduce numerical
two-body relaxation effects, i.e., to make the simulations as collisionless
as possible. However, one still wants to model the density distribution
with sufficiently high spatial resolution, so a trade-off between resolution
and acceptable two-body relaxation has to be made. With this in mind,
previous experience (van Kampen 1995) shows that a Plummer softening
parameter of 40 kpc is a good choice for the dark matter particles.
This corresponds to $1/25$th of the initial mean particle separation,
which is well within the range of values argued for by various other
users of treecodes (e.g.\ \citeNP{bouchet88}; \citeNP{barnes89};
\citeNP{hernquist90b}; \citeNP{dyer93}). The softening of the {\it galaxy
particles} depends on the half-mass radius of the original
simulation group that was `transformed' into a galaxy.
Typical values fall in the range 20--50 kpc (see van Kampen 1996),
which are also well within the range usually found for treecodes.
However, galaxy particles can have masses up to two hundred times
larger than that of the dark matter particles, which means that an
interaction between a galaxy and a dark matter particle can
cause a significant change in the velocity of the latter, despite
the softening. On the other hand, the dark matter particles outnumber
the galaxies 50 to 1. For the models studied in this paper the collective
contribution of galaxy particles to the relaxation of dark matter
particles is up to five times that of the dark matter particles
themselves (van Kampen 1996). However, for our models the corresponding
relaxation time is still at least 6 times $t_0$, the present age of the
universe (30$t_0$ if we had just dark matter particles). For the
chosen value of $H_0$ here, ${t_0}\,\sim\,{1\times {10^{10}}}$ yr.
So while simultaneously retaining the collisionless nature of the
simulation, we do need to bear in mind the differing contributions of
the two components to the deflection of particle orbits. 

\section{Constructing the DF for galaxy cluster models}

\begin{figure}
\psfig{figure=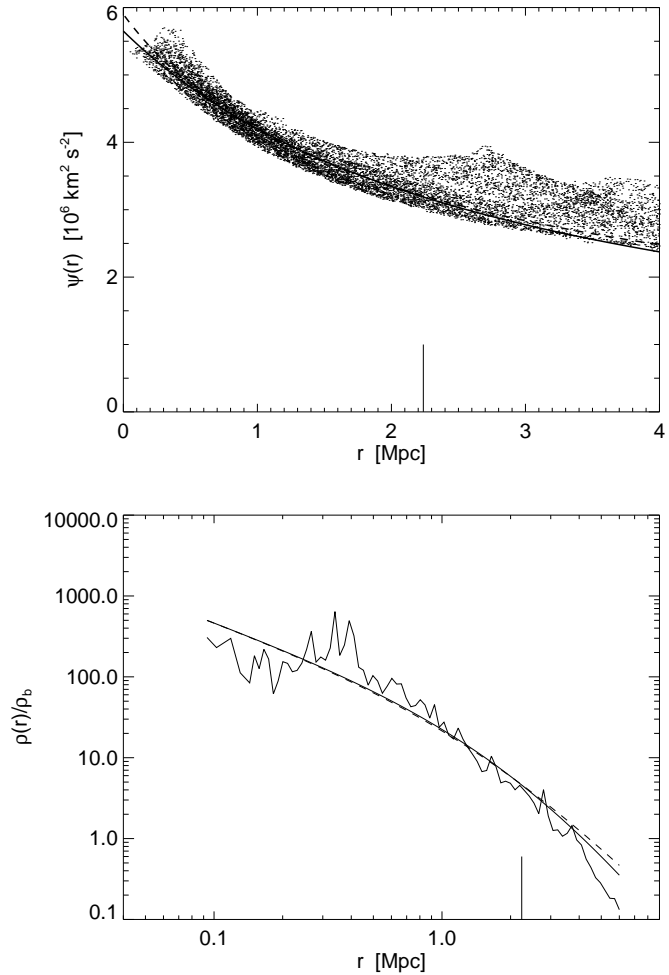,width=9cm}
\caption{The density profile and potential for Model 20
for the snap-shot taken at $z=0$ are plotted above. 
In the top panel ({\bf a}) the points are the exact
potential from the simulation and overplotted are the fit to the
Hernquist profile (solid curve)[with fit parameters: $\Psi_0$ =
5.6 $\times{10^6}$ km$^2$ s$^{-2}$; s = 2.89 Mpc] and the NFW profile 
(dashed curve) [with fit parameters $\Psi_{\rm NFW}$ = 6.44 
$\times{10^6}$ km$^2$ s$^{-2}$; s = 1.09 Mpc]. The lower panel ({\bf
  b}) shows the spherically averaged density profile overlaid with the Hernquist 
model (solid curve) [with fit parameters: $\rho_0$ = 50.70 $\rho_b$; 
s = 3.18 Mpc] and NFW model (dashed curve) [with fit parameters: 
$\rho_0$ = 51.33 $\rho_b$; s = 1.82 Mpc].
In both panels the large tick mark on the x-axis marks the value of
$r_{180}$ = $r_{\rm max}$ = 2.24 Mpc.}
\end{figure}
Since the positions, velocities and energies of all the particles that
constitute a cluster model grown in a cosmological N-body simulation
are known, the distributions $h({\cal E})$, $g({\cal E})$ and
$f({\cal E})$ can be evaluated. In this section we describe the exact
procedure followed to construct the DF for such a system.

For any given snapshot we first compute the centre of mass enclosed
by clumps within a 12 Mpc shell and then recompute this within 2
Mpc around the most massive clump. Any systematic velocity with
respect to the centre of mass is subtracted, which is equivalent to 
ensuring that the main clump 
defines the origin of phase-space. This in turn defines 
$\Psi (\bmath r)$ and $\bmath v$, and thus $\cal E$ through eq.~(4).
It is then straightforward to compute the differential energy
distribution $h({\cal E})$ as a histogram of the distribution of
mass in bins of energy.

Whereas no assumptions regarding the geometry of the configuration 
enter the definition of $h({\cal E})$, our aim of computing 
$f({\cal E})$ necessarily incorporates the approximation of spherical
symmetry. 
This occurs
during the computation of $g({\cal E})$ wherein although the exact 
potential for the particles is known precisely from the simulation, a
spherically averaged value is used. The potential for Model 20 ($z=0$) is shown in
Fig.~3a. Rather than simply averaging the observed potential
in each radial bin we fit an analytical form to the data points.
Overplotted are fits to the Hernquist potential and the NFW potential.
Both yield acceptable fits for the spherically averaged
potential and density profile (Fig.~3b) within the virial radius.
Note here from Fig.~3a that at the virial radius the potential is
roughly half of its central value, which is still deep inside 
the potential well. Therefore matter from the infall region 
does not significantly influence the dynamics within the virialized region. 

Examining several morphologically
distinct cluster models we find that the quality of the fit to these 
analytic models depends crucially on the degree of substructure in the
cluster which shows up as the spread in $\Psi(r)$ in Fig.~3a. The
fits get progressively worse for clumpy models since
for these the centre of the potential is not well-determined and
the shape departs maximally from sphericity even within the virial
radius. 

The advantage of using the
Hernquist potential fit is that the modified density of states,
\begin{eqnarray}
g({\cal E})\,=\,{16 {\pi^2}}{\int_0^{r_m({\cal E})}}
{\sqrt{2\Bigl({\Psi_0 \over {1+{r \over s}}} - {\cal E}\Bigr)}} r^2 dr,
\end{eqnarray}
can be computed analytically (see Appendix A). The disadvantage
of using such a fit to a given restricted form is in fact rather
small since a much more severe limitation is our initial assumption 
of spherical symmetry. For these reasons we have used the Hernquist
fit (\citeNP{lacey96}) in our computations of the density of states. However, we must
emphasize here that the entire procedure is independent of
the precise choice of functional form for the fitted potential 
and we only use the Hernquist fit to the potential
for convenience and because it does provide an elegant analytic 
form for $g({\cal E})$. 

\begin{figure}
\vspace{0.5cm}
\psfig{figure=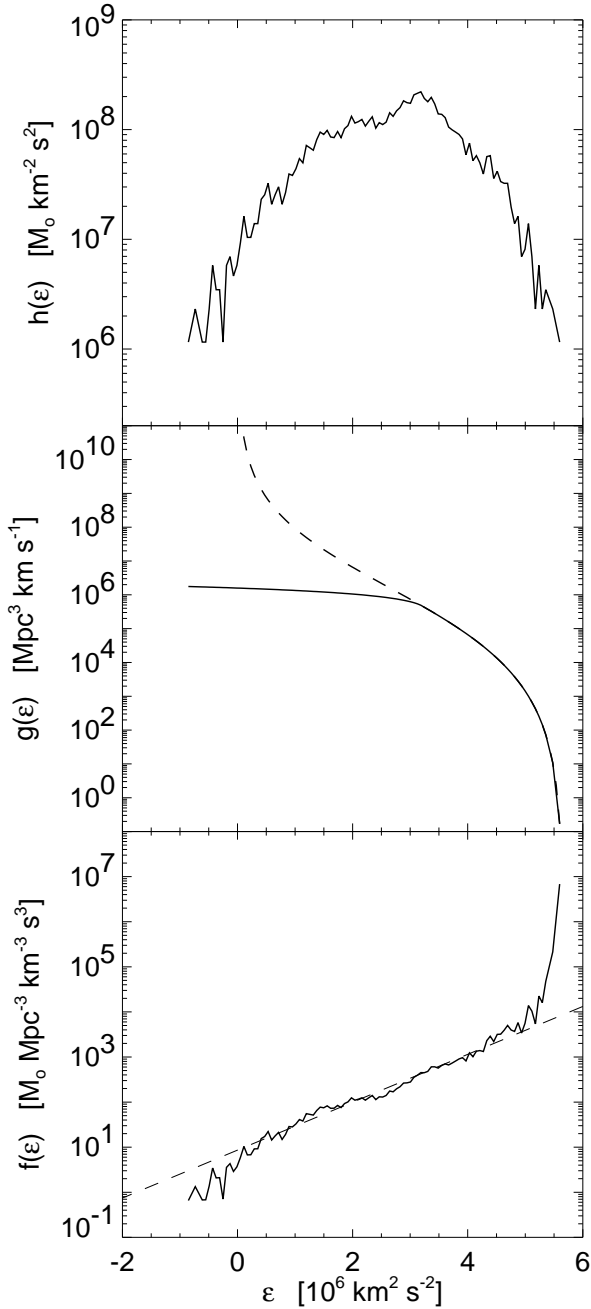,width=8.6cm}
\vspace{1.0cm}
\caption{The computed quantities for a slice of Model 20 ($\Omega_0=1$)
at $z=0$.
The top panel shows the differential energy distribution
$h(\cal E)$ for particles inside the virial radius, plotted as a histogram in energy bins.
The middle panel shows the density of states in each energy hypersurface
$g(\cal E)$. The dashed curve is the usual equilibrium definition of
$g(\cal E)$ (eq.~8) computed analytically using a Hernquist fit to the
potential, which diverges at ${\cal E}=0$ (cf.\ Fig.~3). The solid curve uses
equation (25), i.e., with the devised truncation (19) prescription
presented in Fig.~1. Finally, the bottom panel shows the distribution
function $f(\cal E)$ obtained by dividing $h(\cal E)$ (top panel) with
the modified $g(\cal E)$ (solid curve in middle panel). The overplotted
dashed line is the best particle-number weighted exponential fit to
the curve.}
\end{figure}

The distribution function $f({\cal E})$ is then constructed using the
definition in equation (9). For the snapshot at $z\,=\,0$ for Model 20,
the computed $h(\cal E)$, $g(\cal E)$ and $f(\cal E)$ are shown in Fig.\ 4. 
\mbox{From} this figure we see that $h({\cal E})$ (top panel) has a broad
distribution and has a small fraction of positive energy 
particles. The effect of our prescription for truncating the system
in the computation of $g({\cal E})$ is shown in the middle panel,
where the solid curve corresponds
to the restricted $g({\cal E})$ and the dashed curve shows the
asymptotic behavior (divergent) in the vicinity of ${\cal E}\,=\,0$.
In the bottom panel, we see that the computed $f({\cal E})$ is
remarkably well-fit by an exponential function of the form
$f({\cal E})\,=\,{f_0}{e^{-{\cal E}/{{\sigma_{\rm fit}}^2}}}$
(the overplotted dashed curve), in the region dominated by the
tightly bound particles, except in the deep potential
region. We discuss this effect further in Section 5.2.

Defining the dimensionless central potential,
\begin{eqnarray}
W_0\,=\,{{\Psi_0} \over {\sigma_{\rm fit}}^2},
\end{eqnarray}
the equivalent {\it r.m.s.}\ 1D velocity dispersion
$\sigma_{\rm 1D}(r)$ can be computed analytically given $\sigma_{\rm
  fit}$ and $\Psi_0$ by comparing with
the King model (the lowered exponential, see \citeN{b&t87} for details) which also has an 
exponential $f({\cal E})$,
\begin{eqnarray}
\sigma_f^2\,\equiv\,{1 \over 3}\left<{v^2}\right>\,
=\,1 - {8\over{15\sqrt{\pi}}}\delta
{W_0^{-5 / 2}} \,{\sigma_{\rm fit}^2},
\end{eqnarray}  
with
\begin{eqnarray}
\delta\,=\,{\exp (W_0)}\,{{\rm erf} (\sqrt{W_0})}\,-\,{\sqrt{{4{W_0}} \over 
{\pi}}}{\Bigl(1 + {{2{W_0}} \over 3}\Bigr)}.
\end{eqnarray}  
The velocity dispersion computed from the fit to the 
constructed isotropic distribution function $f({\cal E})$
agrees well (to within $5\%$) with the mean
1D velocity dispersion inside $r_{180}$ computed directly from the simulation.

The DF can also be compared to a characteristic phase-space density 
\begin{eqnarray}
{\bar f} \equiv {3 \over 8 \pi^{5/2}} {M(r_{\rm max}) \over r_{\rm max}^3 
\sigma_{\rm 1D}^3},
\end{eqnarray}
motivated by the expression for the central phase-space density of the
isothermal sphere, $f_c=(2\pi)^{-3/2}\rho_c\sigma^{-3}$.
For Model 20, which has $M(r_{\rm max})=5.87\times10^{14} M_\odot$,
$r_{\rm max}=2.24$ Mpc and $\sigma_{\rm 1D}=970$ km s$^{-1}$,
we find ${\bar f}\approx 1200$ M$_\odot$ Mpc$^{-3}$ km$^{-3}$ s$^{3}$. 
This fits in nicely with the distribution function $f({\cal E})$ as 
plotted in Fig.\ 4.
These comparisons clearly demonstrate that the entire procedure
produces sensible results, both quantitatively and qualitatively.

To compute the anisotropic distribution function $f({\cal E},L)$ we
use the definitions in equations (12), (23) and (24), once again with
${r_{\rm max}}=r_{180}$, the expected virial radius.
\begin{figure}
\psfig{figure=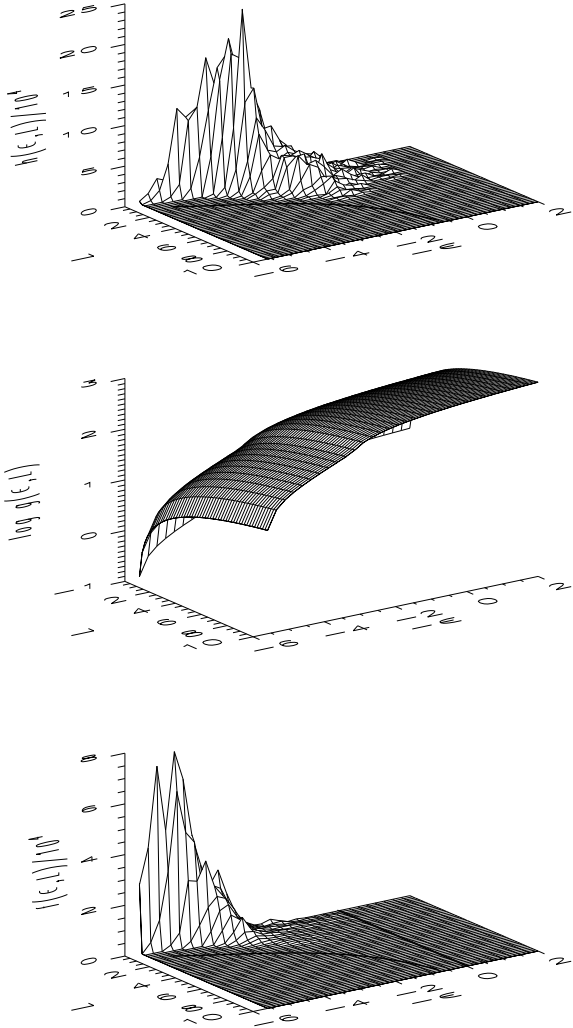,width=9cm}
\caption{The anisotropic distribution function. The figures show
(from the top to bottom) the differential distribution
$h({\cal E}, L)$, the modified density of states $g({\cal E}, L)$ and 
the constructed distribution function $f({\cal E}, L)$ for Model 20. 
Here $h({\cal E}, L)$ is in units of M$_\odot$
Mpc$^{-1}$\,km$^{-3}$\,s$^{3}$, $g({\cal E}, L)$ in units of M$_\odot$
Mpc$^{2}$ and $f({\cal E}, L)$ in units of M$_\odot$ Mpc$^{-3}$
km$^{-3}$ s$^{3}$. Here ${\cal E}$ is in units of ${10^6}$\,km$^{2}$
s$^{-2}$ and $L$ in units of Mpc km s$^{-1}$.}
\vspace{2.0cm}
\end{figure}
The surface plots for $h({\cal E}, L)$, $g({\cal E}, L)$ and
$f({\cal E}, L)$ are shown in Fig.~5. The $h({\cal E}, L)$
distribution has a coherent peak at low values of angular momentum
and energy. Unlike \citeN{voglis}, who analysed galaxies, we do
not find a differentiated core and halo population for a galaxy cluster
on equivalent scales, although there exists a small fraction of positive 
energy particles. The anisotropic DF is sharply peaked and 
falls off to zero for large ${\cal E}$, similar to $f({\cal E})$.
To conclude this section, we have demonstrated that DFs can be
constructed self-consistently for N-body model clusters 
and that sensible results are obtained.

\section{Results}

\begin{figure}
\psfig{figure=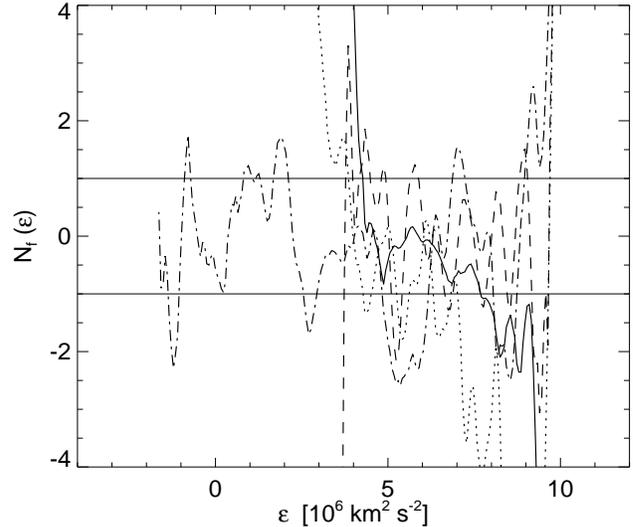,width=9cm}
\caption{The fractional change in $f({\cal E})$, $N_f({\cal E})$
during a typical dynamical time for the four clusters computed at
$z=0$ are plotted above. The solid curve corresponds to Model 7 (evolved in an
$\Omega_0\,=\,0.2$ universe), the dotted curve to
Model 7, the dashed curve to Model 20 and the dot-dashed curve to 
Model 41 (all three evolved in an $\Omega_0\,=\,1.0$ model). The
horizontal lines correspond to $N_f = 1$ and $-1$.}
\end{figure}

In this section we analyse the dynamical states of the model 
clusters (restricted to the virialized central region by construction)
using the time evolution of the DF in conjunction with other
physically interesting quantities. 
Before embarking on a discussion of the various distribution
functions, it is instructive to briefly
outline the evolution as seen in the N-body simulations for the
studied models shown in Fig.~2.

Model 7 evolving in the $\Omega_0\,=\,0.2$ universe, is relatively
isolated and its evolution is being traced during a passively evolving
phase. Model 7 ($\Omega_0\,=\,1$) has a
significant amount of substructure and undergoes a violent merging event
at the present epoch, while Model 41, which is a
significantly more massive cluster, is being followed during one of
its more quiet evolutionary stages. 

The three $\Omega_0=1$ models are typical for the standard CDM scenario
in the sense that a cluster does not form from a single
collapse but rather from a heterogenous collapse, typically from two or
more diffuse sub-clumps that grow individually and merge to form the final
cluster. Before and after these mergers the cluster grows by accreting
material along several filaments, which join
at the position of the cluster. This `secondary' infall proceeds in a
somewhat lumpy fashion. The complex aggregation process implies 
that present-day cluster cores are often obscured by surrounding
clumps. The fact that these clumps are falling towards the cluster makes
it hard to distinguish them when observing along the line of sight in
redshift space. 
The $\Omega_0=0.2$ cluster model is also representative for its cosmology,
in the sense that clusters form early in this scenario, show only
little secondary infall, and few merging clumps.
We therefore expect this model to be relatively close to equilibrium,
even though small clumps are still falling in and the dynamical
timescale is about a sixth of a Hubble time.

In the following subsections we briefly discuss what can be learnt
from this new DF approach. The emphasis will be on demonstrating the
usefulness and reliablitity of our construction of distribution
functions, using the N-body models to illustrate the results.
Detailed discussions of the dynamics of cluster formation and the
cosmological implications using this approach are deferred to later
papers.

\subsection{Time scales and quasi-equilibrium}

As anticipated in Sections 2.1 and 2.3, it is evident from the
evolution plots in Figs.~7, 8 and 9, that in general, even over
relatively short time scales (compared to the Hubble time),
the dynamical state changes perceptibly, thus validating the
quasi-equilibrium approach of treating the system as one evolving
over two characteristic but disparate time-scales.

To test the assumption that clusters are indeed in a quasi-equilibrium state
at least in the interim between major merger events, we calculate the fractional
change in $f({\cal E})$ during one dynamical time
$t_{\rm dyn}\equiv{\sqrt{3\pi/16G\rho}}$ (\citeNP{b&t87}), which is equal
to $0.25 t_0$ (for the flat model), as $\rho=\rho_{180}=180\rho_b$ for the
virialized region under study (by definition).
At $z=0$ we compute the dimensionless quantity $N_f({\cal E})$ defined to be
a measure of the stationarity,
\begin{eqnarray}
{N_f}({\cal E})\,\equiv\,{t_{\rm dyn}}\,({\frac {{\rm d}f}{{\rm
      d}t}}\cdot{\frac{1}{f}}),
\end{eqnarray}
plotted in Fig. 6.
In order to get a quantitative measure for the stationarity of
$f({\cal E})$, we calculate the $h({\cal E})$-weighted averages of
$N_f({\cal E})$ for all four models and hence define the
`stationarity index' $n_f$ as, 
\begin{eqnarray}
{n_f}\,\equiv\,{{\int{|{N_f}({\cal E})|{h({\cal
        E})\,d{\cal E}} }\over M}},
\end{eqnarray}
where $M$ is the total mass enclosed within $r_{180}$ for the system. 
The values of
$n_f$ that we find are: 0.32, 3.3, 3.7, and 0.96
for Model 7 ($\Omega_0=0.2$), 7 ($\Omega_0=1$), 20, and 41 respectively.
We interpret this as follows: only Model 7 for the $\Omega_0=0.2$ case
can really be considered to be in quasi-equilibrium, with clear skewing evolution of
$f({\cal E})$, while the same model within an $\Omega_0=1$ universe,
where it is undergoing a major merger, and Model 20 are furthest from an
equilibrium state. Model 41 is a `critical' case: there certainly is
on-going secondary infall, but the cluster is almost massive enough
for the virialized region to remain in quasi-equilibrium.
Therefore, we roughly delineate three distinct regimes on the basis of the
values of the `stationarity' index $n_f$: ${n_f}\,<\,1$ corresponds to
the regime where the quasi-equilibrium description is 
an adequate one, ${n_f}\,=\,1$ to the critical case and
${n_f}\,>\,1$ to the case when the system has departed from
quasi-equilibrium while recuperating from a merger.   

\subsection{Signatures of substructure}

As mentioned in Section~4 there appears to be a central peak in
$f({\cal E})$ for most of the models. The peak is particularly marked 
when there is significant substructure
inside the studied region. We therefore associate this peak with 
infalling sub-clumps outside the cluster core which already have a 
deep `local' central
potential, but contain an insignificant fraction of the mass of the
system. Indeed, the effect is not seen in
$h({\cal E})$ confirming the above picture. The central peaks in
$f({\cal E})$ are thus partially an artifact of the assumption of 
spherical symmetry of the potential, but are also a signature of 
transient substructure. Once the clumps are incorporated or fall in to
the centre of the cluster, the peak disappears.

The radial profile for the velocity anisotropy parameter,
\begin{equation}
\beta (r)\equiv 1-{\sigma_\theta^2(r)\over\sigma_r^2(r)},
\end{equation}
is correlated to the presence of substructure (see Fig.\ 2). In the 
absence of significant sub-clustering the velocity distribution
is found to be roughly 
isotropic but erratic ($\beta \approx 0$) and for quiescent evolution $\beta>0$,
indicating the presence of primarily radial orbits, which are most
likely to originate in the infall region. 

The presence of substructure in the cluster models being progressively
subsumed into the main clump can be seen in the (significant)
evolution of the 1D velocity dispersion profiles as well. For example,
consequent to a merger the mean value increases as shown in the
top left panel of Fig.~8. Furthermore, the overall shape seems to be
well correlated with substructure and secondary infall. This promises
to be one of the important applications of the DF formalism.

\subsection{Evolution in the density profiles}

Comparatively little systematic evolution is seen in the density
profiles, aside from the progressive worsening of the fits to the
analytic form occuring due to the presence of sub-clumps causing
significant departures from sphericity, and small changes in the
value of the central density. Density profiles seem to be set up 
rather early in the evolutionary history of the cluster. 

\subsection{The central potential and degree of central condensation}

\begin{figure*}
\psfig{figure=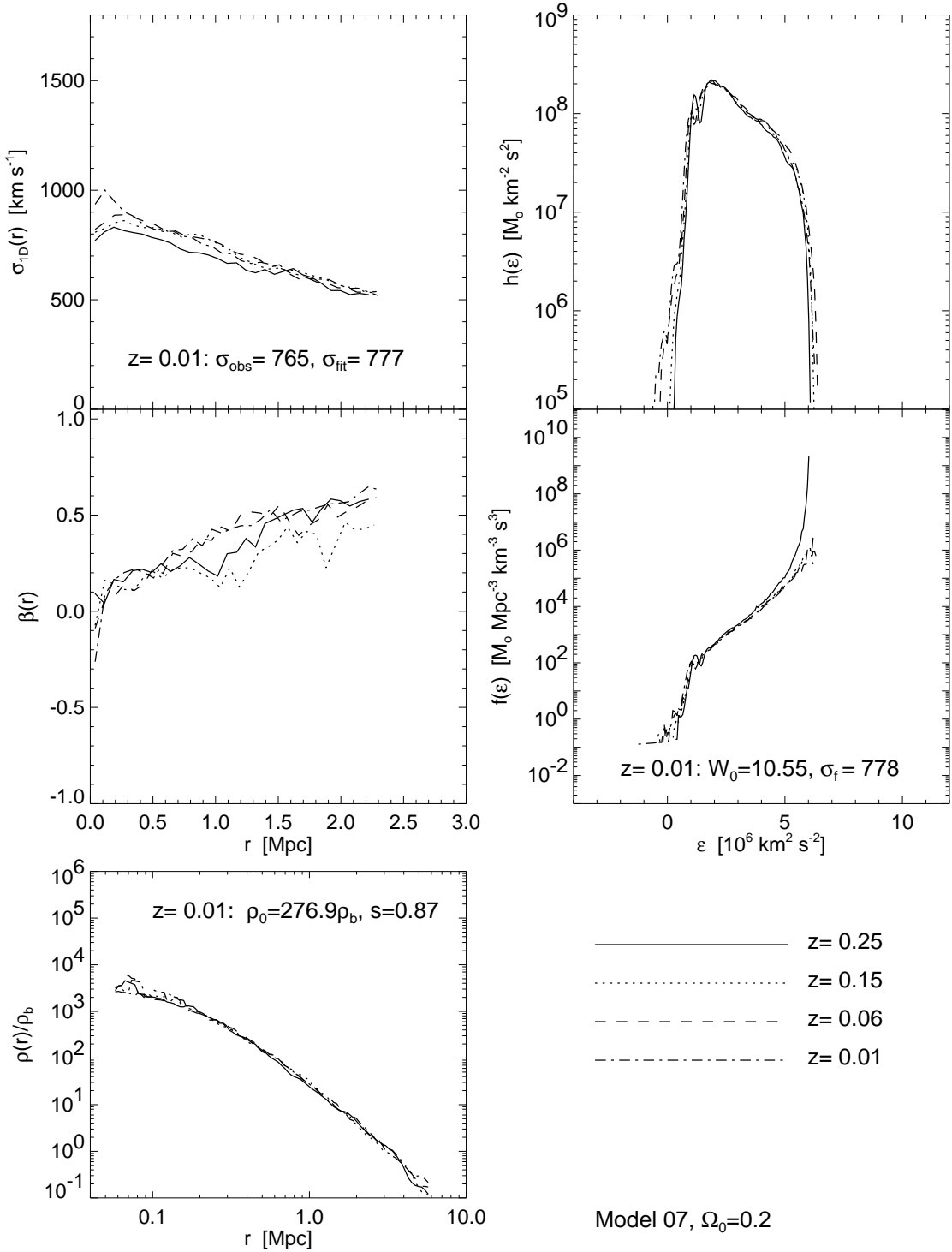,width=15cm}
\caption{The evolution of computed quantities for a slice of Model 7
($\Omega_0=0.2$) for four redshifts chosen to correspond to constant 
time intervals. As a function of physical radius,
the left panel shows the measured one-dimensional velocity dispersion
$\sigma_{\rm 1D}(r)$ (top), the anisotropy parameter $\beta (r)$
(as defined in eq.~32) (middle), and the density $\rho (r)$ relative to the background 
density $\rho_b$ (bottom). In the right panel
the differential energy distribution $h(\cal E)$ (top) and the computed 
isotropic distribution function  $f(\cal E)$ (middle) are plotted as 
a function of the binding energy per unit mass $\cal E$.}
\end{figure*}

\begin{figure*}
\psfig{figure=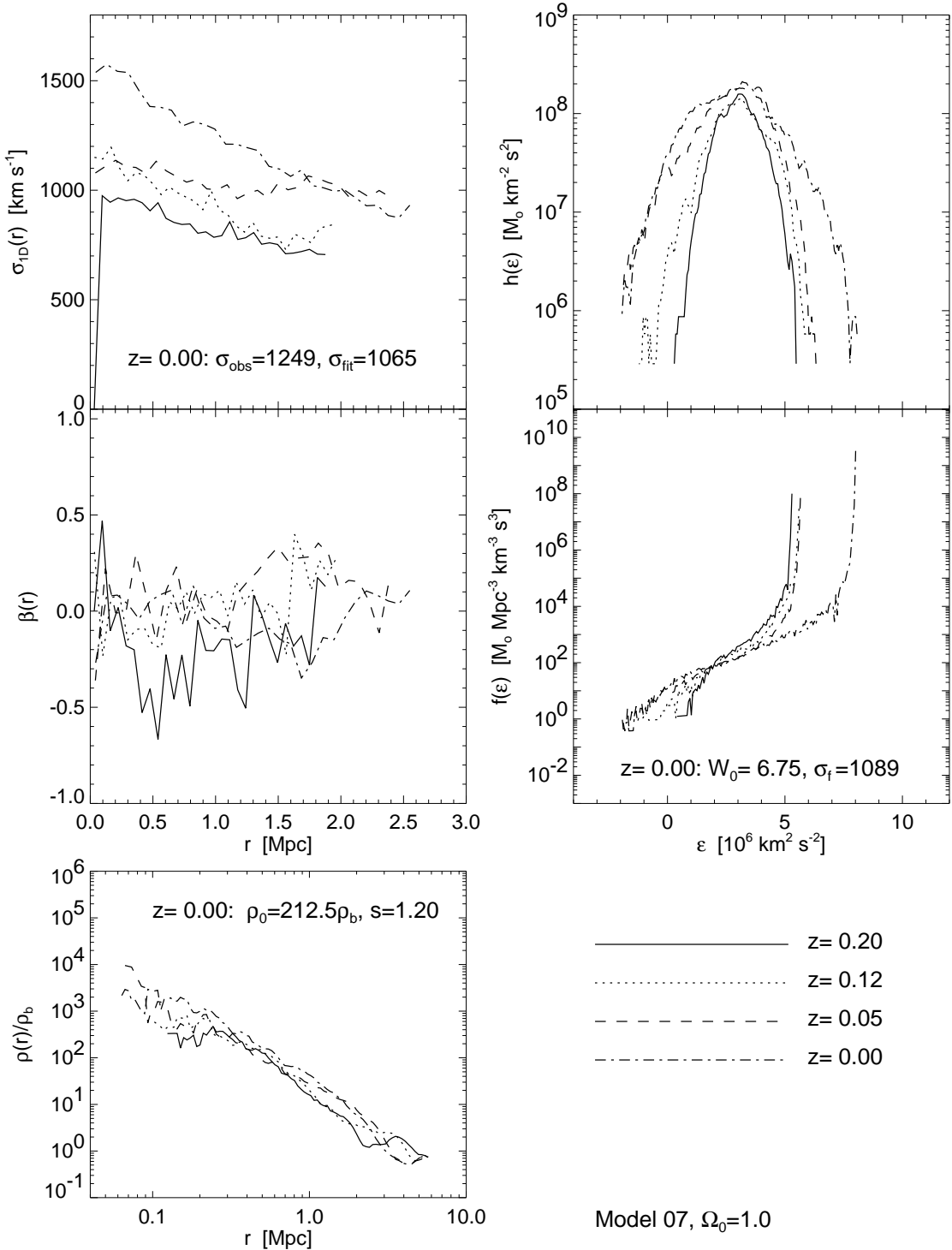,width=15cm}
\caption{Same as Fig.~7 for Model 7 within an $\Omega_0=1$ Universe.}
\end{figure*}

\begin{figure*}
\psfig{figure=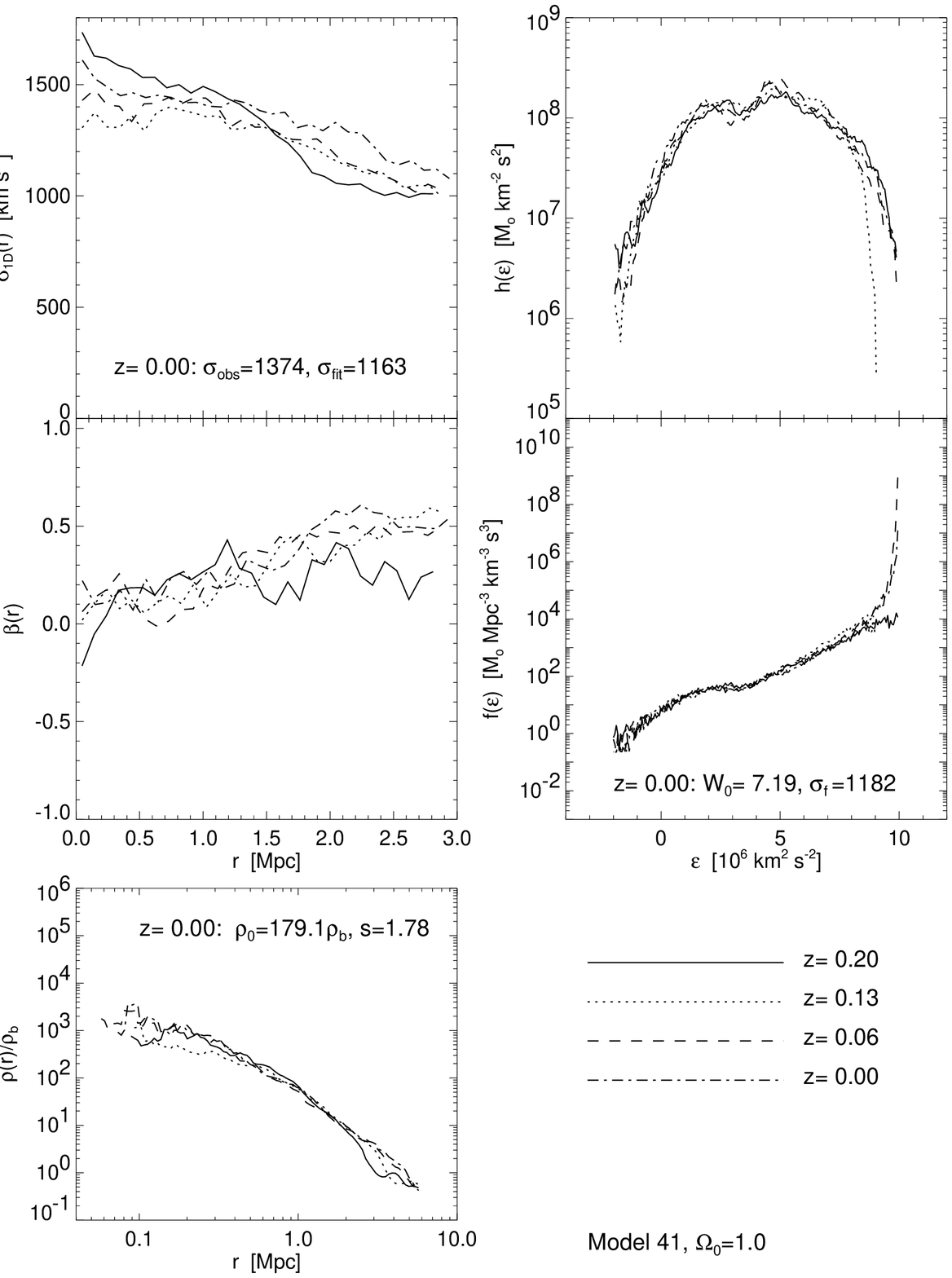,width=15cm}
\caption{Same as Fig.~7 for Model 41 ($\Omega_0=1$).}
\end{figure*}

The dimensionless depth of the potential, $W_0$ as defined in equation (26),
is a measure of the central concentration of the system. For the models
studied, we find $W_0\approx 7$ for the flat models and $W_0\approx 10.5$
for the open model. This is comparable to what is
found for elliptical galaxies, where $8\,\la\,{W_0}\,\la\,11$
(\citeNP{hjorth95}), and uncollapsed globular clusters for which 
${W_0}\,\la\,10$ (eg.\ Binney \& Tremaine 1987). 
Clusters evolving in an Einstein-de Sitter universe have
comparatively shallower central potentials since the system is 
constantly disturbed by infalling sub-clumps and is hence 
unable to build up a concentrated core in the elapsed time.
By contrast, a cluster evolving in an open universe is
comparatively more homogeneous and relatively undisturbed. It evolves
quiescently essentially due to the fact that dark matter particles are
allowed to plunge down into the centre on radial orbits (as evidenced
by $\beta\,>\,0$ seen in the middle left panel of Fig.~7)
hence building up to high central densities.

The observed trends in the central concentration can also be
potentially related to features of the initial conditions by 
estimating the characteristic phase-space density $\bar f$. The typical
maximum phase-space density is about 100 times larger than this
characteristic value, similar to what has been found for elliptical 
galaxies. Since the maximum phase-space density cannot increase during 
collisionless evolution for an isolated system
(\citeNP{tremaine86}), the computed value of $\bar f$  
allows us to understand the relative efficiency
required of any relaxation process, given characteristic
time-scales of mediation, in order to produce the present
configuration from the inhomogeneous initial cosmological 
conditions from which a cluster first assembles. 

\subsection{Energy distribution}

There is a clear difference between the differential energy distributions, 
$h({\cal E})$, of the open and flat models. The energy distribution
for the open model (Fig.~7) is very similar to that of elliptical
galaxies where a `negative temperature' exponential function,
cut-off close to the escape energy (${\cal E}=0$), provides a
good fit the $h({\cal E})$ (Binney 1982), i.e.,
most of the particles are loosely bound. By contrast, the energy 
distribution of the flat models are better approximated by
gaussians with a tail of positive-energy particles.
The advantage of our adapted DF formalism is that it allows the inclusion
of positive energy particles as well. For the galaxy cluster models
evolved in a flat universe we find that the particles with 
positive energies that are located preferentially in the centre of 
the cluster, despite the deep potential well. Energy exchange due to 
violent merger events as seen in Model 7 for $\Omega_0 = 1.0$, for instance,
are a potential source for generating these unbound particles,
as we clearly see a significant fraction of them just before and
during a merger. We also detect simultaneously a surge in
the maximum value of ${\cal E}$, indicating significant energy diffusion.
We hope to address this issue in the context of examining the
relaxation process in more detail in a future paper.

\section{SUMMARY AND CONCLUSIONS}

In this paper we have applied an adapted version of the
distribution function formalism of equilibrium stellar dynamics to
numerical models of clusters of galaxies. Clusters are generally not
in equilibrium as on-going secondary infall and the relatively long
dynamical time-scale prevent them from reaching
an equilibrium configuration. However, as evidenced from observations
of clusters, this deviation 
from equilibrium is not expected to be too severe, at least within the 
finite central virialized region, and so a quasi-equilibrium
description can be formulated enabling the construction of the
distribution function for the N-body cluster models. 

Isotropic $f({\cal E})$ and anisotropic $f({\cal E},L)$ distribution
functions can be computed using the 
differential energy distribution and the density of available states
in phase space for the system. The accessible phase space however,
needs to be restricted in a physically meaningful and self-consistent
fashion. We invoke the finite elapsed time since formation and 
space available as providing a natural way to limit phase space. 
We argue that this prescription can be suitably applied when the
analysis is restricted to the virialized region, 
which is defined by $r\,\leq\,r_{180}$. This is motivated by the fact that the
particles that once enter and get bound within this region are not likely
to leave. Even particles that do become unbound due to energy-exchange
processes are likely to lose energy fairly rapidly in the dense
cluster environment on very short time-scales, i.e., their occupancy of
states in phase space is automatically limited by the fact that particles are
expected to change energy again within a relatively short time. It is
precisely the existence of these two disparate time scales (the long
dynamical time scale and the short characteristic time scale for
energy exchange) which implies that
the evolution of the system can be effectively decoupled 
into `fast' and `slow' components and hence allows the description of 
quasi-equilibrium states somewhat within the equilibrium formalism.
In the anisotropic case it is additionally assumed that $g({\cal
  E},L)$ is separable.

We show that this construction can be successfully accomplished for
N-body cluster models taken from a large sample of models for the
standard CDM cosmological scenario, and that the constructed DFs do
show similarities to equilibrium ones. The velocity dispersion as
obtained from an exponential fit to $f({\cal E})$ agrees well
with that obtained directly from the particle distribution in
the N-body simulation (agrees to within 5 \% typically). This is not only
true for specific redshift slices, but also during the course of
evolution. The only exceptions are during strong merger events, whence
the difference can be a few hundred km s$^{-1}$.
The characteristic phase-space density, calculated from global 
properties of the particle distribution in the simulation and from the
computed DF also agree well.

Furthermore, we show by following the time variation
of the DFs that they are good tracers of dynamical events like 
cluster mergers and infall of clumps. There is notable evolution with
redshift for most models, indicating on-going dynamical evolution. 
The exception is the cluster model simulated in an 
$\Omega_0=0.2$ CDM universe, which shows only mild evolution and can
be said to be in quasi-equilibrium. It is also the only model to show
a linear rise of the velocity anisotropy parameter $\beta$ from zero
in the centre of the cluster to about 0.5 at $r_{180}$. This
cluster bears the closest resemblance to an elliptical 
galaxy, further demonstrated by the shape of $h({\cal E})$ and
the degree of central concentration 
which is well within the range typically estimated for ellipticals. 
We also find that the density
profiles within the virialized region show practically no evolution
and seem to be set up quite early in the dynamical history.

As to the possibilities of using the constructed DF to describe the
dynamical state of a galaxy cluster, it can clearly distinguish the 
merging clusters from the quiet ones, as well as the ones dominated by
secondary infall. The evolution of the DF, even over a relatively
short time, reveals the on-going kinematic activity. The comparison of
global properties like the velocity dispersion as obtained from a fit
to $f({\cal E})$ to those calculated directly from the simulation
provide a new indicator to establish whether the cluster is close to 
equilibrium.

In a recent study of the evolution 
of cluster-scale dark halos, \citeN{tormen96} also characterize the halo 
formation process as being composed of alternating merging and 
relaxation phases. The general picture that emerges from our analysis
is one of clusters evolving quiescently with on-going
collective energy exchange (which we define as quasi-equilibrium
states) punctuated by violent merging phases followed by fairly brief,
but yet well-defined recovery phases. During the passively evolving 
phases, we find that the DF $f({\cal E})$ is well-approximated by an 
exponential and significant deviations occur only during the merging 
and recouping phases. In this treatment, we have not quantified the frequency
or nature of energy exchange processes, aside from the observation
that they do indeed occur. Besides possibly allowing a better
understanding of the physics of energy exchange, this approach also 
potentially affords new discriminants of the underlying
cosmological model.

In conclusion, we have shown that distribution functions can be
successfully constructed for the dark matter component of
evolving, non-isolated clusters of galaxies grown in N-body simulations.
The necessary machinery is thus in place for future work in which we
hope to address the use of distribution functions as a new probe of 
cluster dynamics and evolution, fundamental cosmological parameters,
and the possible origin of `universal' cluster density profiles.

\section{ACKNOWLEDGMENTS}

PN acknowledges Martin Rees and Donald Lynden-Bell for encouragement 
and insightful discussions during the course of this work. PN and JH thank 
John Peacock and the ROE for hospitality where a portion of this work was done.
PN acknowledges financial support from the Isaac Newton Studentship and Trinity
College, Cambridge; JH from the Danish Natural Science Research Council
(SNF) and EvK the receipt of an EC HCM Research Fellowship.
We acknowledge useful discussions with Bernard Jones, Steinn
Sigurdsson, Simon White and Tim de Zeeuw.
\onecolumn

\appendix
\section{Truncated densities of state for various potentials}

For the \citeN{hernquist90a} potential used in this paper,
\begin{eqnarray}
\Psi(r)\,=\,{{\Psi_{0}} \over 1 + {r \over s}}.
\end{eqnarray}
where $s$ is a characteristic scale radius, the truncated density of
states can be expressed analytically as,
\begin{eqnarray}
{g({\cal E}) \over 16 \pi^2} 
\,&=&\, 
-{\sqrt{\Psi_0-{\cal E}}\, ( 8\,{{\cal E}^2} + 10\,{\cal E}\,{\Psi_{0}} - 3\,
{{{\Psi_{0}}}^2}  ) \,{s^3} \over {12\,{\sqrt{2}}\,{{\cal E}^2}}} \\ \nonumber 
&+& 
{{{A \left ( 8 {\cal E}^2 (r_m^2+s^2-r_m s)+10 {\cal E} 
\Psi_0 s^2 - 3 \Psi_0^2 s^2 - 2 {\cal E} \Psi_0 r_m s \right )
(r_m+s) }\over {12\,{\sqrt{2}}\,{{\cal E}^2}}}} \\ \nonumber 
&+& 
{{{\Psi_{0}}\,\left( 8\,{{\cal E}^2} - 4\,{\cal E}\,{\Psi_{0}} +
{{{\Psi_{0}}}^2} \right) \, {s^3}\,\over{8\,{\sqrt{2}\,{{\cal E}^{5/2}}}}}}
\left [\,{{\tan^{-1}} \left ({\Psi_0-2\,{\cal E}\over 
{2\,{\sqrt{{\cal E}}}\,{\sqrt{\Psi_0-{\cal E} }}}}\right )} - {{\tan^{-1}} 
\left ({{A\left( 2\,{\cal E}\, {r_{m}} + 2\,{\cal E}\,s - {\Psi_{0}}\,s \right) 
}\over {2\,{\sqrt{{\cal E}}}\,\left( {\cal E}\,{r_{m}} + {\cal E}\,s - 
{\Psi_{0}}\,s \right) }}\right )} \right ]
\end{eqnarray}
for ${\cal E} > 0$ and
\begin{eqnarray}
{g({\cal E}) \over 16 \pi^2} 
\,&=&\, 
-{\sqrt{\Psi_0-{\cal E}}\, ( 8\,{{\cal E}^2} + 10\,{\cal E}\,{\Psi_{0}} - 3\,
{{{\Psi_{0}}}^2}  ) \,{s^3} \over {12\,{\sqrt{2}}\,{{\cal E}^2}}} \\ \nonumber 
&+& 
{{{A \left ( 8 {\cal E}^2 (r_m^2+s^2-r_m s)+10 {\cal E} 
\Psi_0 s^2 - 3 \Psi_0^2 s^2 - 2 {\cal E} \Psi_0 r_m s \right )
(r_m+s) }\over {12\,{\sqrt{2}}\,{{\cal E}^2}}}} \\ \nonumber 
&-& 
{{{\Psi_{0}}\,\left( 8\,{{\cal E}^2} - 4\,{\cal E}\,{\Psi_{0}} + 
{{{\Psi_{0}}}^2} \right) \, {s^3}} \over {8\sqrt{2}\,{(-\,{{\cal E})^{5/2}}}}} 
\left  [\,\,\ln\, \left (2s ( \sqrt{\Psi_0-{\cal E}}\sqrt{-{\cal E}}-{\cal E})
+\Psi_0 s \right)
- 
\ln\, \left (2 (r_m+s)(A\sqrt{-{\cal E}}-{\cal E})+\Psi_0 s
\right )\right ]
\end{eqnarray}
for ${\cal E} < 0$. In these expressions,
\begin{equation}
A=\sqrt{\Psi_0 {s \over r_m+s} - {\cal E}}.
\end{equation}

Alternate popular potentials do not completely reduce to analytic
expressions. For the NFW potential (\citeNP{navarro96}),
\begin{eqnarray}
\Psi (r)\,=\,{{{\Psi_{\rm NFW}}\,s} \over r}\,{\ln (1 + {r \over s})},
\end{eqnarray}
the truncated density of states is
\begin{eqnarray}
{g({\cal E}) \over {16 \pi^2}}\,=\,{{2  \over {\sqrt 5}}} {\int_{0}^{r_m}\,{{dr\,
{r^{5 \over 2}}\,({\cal E}\,-\,{{\Psi_{\rm NFW}}{s \over {r + s}}})} 
\over {\sqrt{{{\Psi_{\rm NFW}}\,s\,{\ln (1 + {r \over s})}}\,-\,{r {\cal
 E}}}}}}\,+\,{{{2\sqrt 2} \over 5}{{r_m}^{5/
2}}{\sqrt{{{\Psi_{\rm NFW}}\,s\,{\ln (1 + {r \over s})}}\,-\,{{r_m}{\cal
  E}}}}}.
\end{eqnarray}

For the Jaffe model (\citeNP{jaffe83}),
\begin{eqnarray}
\Psi(r)\,=\,{\Psi_{J}}\, \ln (1 + {{r_{J}} \over r}),
\end{eqnarray}
the truncated density of states is
\begin{eqnarray}
{{g({\cal E})} \over {16 {\pi^2}}}\,=\,  {{{{\Psi_{J}}\,{r_{J}}\over {3\,{\sqrt{2}}}}}\, \int _{0}^{{{r_
{m}}}}
          dr\,{{{r^2}}\over 
             {( r + {{r_{J}}}) \,
               {\sqrt{{{\Psi_{J}}}\,\ln (1 + {{{{r_{J}}}}\over r})-{\cal E}}}}} 
       }
 + 
  { {{\sqrt{2}}\over 3}\,{{{{r_{m}}}}^3}\,
       {\sqrt{{{\Psi_{J}}}\,\ln (1 + {{{{r_{J}}}}\over {{{r_{m}}}}})-{\cal E}}}}.
\end{eqnarray}

For the singular isothermal sphere (\citeNP{b&t87}),
\begin{eqnarray}
{\Psi(r)}\,=\,{{\Psi_{I}}\,\ln\,({1 \over r})},
\end{eqnarray}
the truncated density of states is
\begin{eqnarray}
{{g({\cal E})} \over {16 {\pi^2}}}\,=\,{{{\Psi_{I}\over {3\,{\sqrt{2}}}}}\, \int _{0}^{{{r_{m}}}}
          dr\,{{{r^2}}\over {{\sqrt{{\Psi_{I}}\,\ln ({1\over r})-{\cal E}}}}}
           } + 
   {{{\sqrt{2}}\over 3}\,{{{{r_{m}}}}^3}\,
       {\sqrt{ {\Psi_{I}}\,\ln ({1\over {{{r_{m}}}}})-{\cal E}}}}.
\end{eqnarray}

\bibliography{mnrasmnemonic,refs}
\bibliographystyle{mnrasv2}
\end{document}